\documentclass[12pt]{article}
\usepackage{a4}
\usepackage{amssymb}
\usepackage{cite}
\newcommand{\be}{\begin{equation}}
\newcommand{\ee}{\end{equation}}
\newcommand{\bea}{\begin{eqnarray}}
\newcommand{\eea}{\end{eqnarray}}

\begin{document}

\def\C{{\mathbb{C}}}
\def\R{{\mathbb{R}}}
\def\s{{\mathbb{S}}}
\def\T{{\mathbb{T}}}
\def\Z{{\mathbb{Z}}}
\def\W{{\mathbb{W}}}
\def\Bbb{\mathbb}
\def\BZ{\Bbb Z} \def\BR{\Bbb R}
\def\BW{\Bbb W}
\def\BM{\Bbb M}
\def\e{\mbox{e}}
\def\BC{\Bbb C} \def\BP{\Bbb P}
\def\CP{\BC\BP}
\def\SU {\mathrm{SU}}
\def\U  {\mathrm{U}}
\def\rmd{\mathrm{d}}
\def\rme{\mathrm{e}}
\def\rmi{\mathrm{i}}

\thispagestyle{empty}
\def\thefootnote{\fnsymbol{footnote}}
\begin{flushright}
{\sf ZMP-HH/08-14}\\
{\sf Hamburger$\;$Beitr\"age$\;$zur$\;$Mathematik$\;$Nr.$\;$318}
     \vskip 2em
\end{flushright}
\vskip 2.0em
\begin{center}\Large
SOME REMARKS ON DEFECTS AND T-DUALITY
\end{center}\vskip 1.5em
\begin{center}
Gor Sarkissian,~
~Christoph Schweigert\,\footnote{\scriptsize
~Email addresses: \\
$~$\hspace*{2.4em}sarkissian@math.uni-hamburg.de, 
schweigert@math.uni-hamburg.de}
\end{center}
\begin{center}
Organisationseinheit Mathematik, \ Universit\"at Hamburg\\
Bereich Algebra und Zahlentheorie\\
Bundesstra\ss e 55, \ D\,--\,20\,146\, Hamburg
\end{center}
\vskip 1.5em
\begin{center} October 2008 \end{center}
\vskip 2em
\begin{abstract} \noindent
The equations of motion for a conformal field theory in the
presence of defect lines can be derived from an action that includes 
contributions from bibranes. For T-dual toroidal compactifications,
they imply a direct relation between Poincar\'e line bundles and the
action of T-duality on boundary conditions. We also exhibit a class
of diagonal defects that induce a shift of the B-field. We finally
study T-dualities for $S^1$-fibrations in the
example of the Wess-Zumino-Witten model on $\mathrm{SU}(2)$
and lens spaces. Using standard techniques from 
D-branes, we derive from algebraic data in rational conformal field
theories geometric structures familiar from Fourier-Mukai transformations.
\end{abstract}

\setcounter{footnote}{0}
\def\thefootnote{\arabic{footnote}}
\newpage

\section{Introduction}

Defect lines in two-dimensional quantum field 
theories are oriented lines separating different quantum 
field theories.
Such defect lines have a rich behaviour: 
\begin{itemize}
\item They can carry defect fields changing the type of the defect, very 
much in the way the type of a boundary condition can change at the insertion of a boundary field. 
\item Several defect lines can merge and split at vertices, leading to
defect junctions.
\item Defect lines with the same conformal field theory on both sides
can start and end at disorder fields. 
\end{itemize}

Defects and dualities of lattice models have been
studied extensively (see e.g.\ \cite{drwa} for a
discussion and references). In fact, the simplest 
realization of a defect line in the Ising model is a line with the
property that all bonds crossing this line have
antiferromagnetic coupling. In conformal field
theories describing the continuum limit of such
theories, defects have only been studied recently
(see e.g.\ \cite{Bachas:2001vj,pezu5,fuRs4}). 

A particularly important subclass of defects are
topological defects: they are characterized by the
fact that  correlation functions do
not depend on the precise position of the
defect line: They do not change   under small continuous changes of
position of the defect line. This topological nature
implies that there is a well-defined notion of
fusion for such defects. It has been
shown \cite{ffrs3,ffrs5} that the fusion
algebra of topological defects contains structural information about 
symmetries and Kramers-Wannier dualities of the theory.
All defects considered in this paper are topological, except
possibly the one implementing a shift on the B-field.

Much information about topological defects is by now available in concrete
classes of models, and in fact the fusion algebra
of topological defects in rational conformal field
theories can be computed explicitly in the TFT
approach to RCFT correlators \cite{fuRs4,fuRs12,scts}. Recently, there has
been some progress for the Lagrangian description
of topological defects in conformal sigma-models, see e.g. 
\cite{Fuchs:2007tx,rusu,Quella:2002ct,Quella:2006de}
for the case of sigma models on group manifolds. 
   In particular, a geometric description of Wess-Zumino terms for
   sigma models with non-trivial B-field background in has been established
   \cite{Fuchs:2007fw,rusu}.
The appropriate framework for
theories with non-trivial background B-field are
hermitian bundle gerbes with connection
\cite{gaw}. In such a situation, defects are
described by bibranes \cite{Fuchs:2007fw}:
submanifolds of the Cartesian product of the
target spaces, equipped with a gerbe bimodule, 
i.e.\ a specific one-morphism of bundle gerbes.
Defect junctions are described \cite{rusu}
by so-called ``inter-bibranes'', which contain as part of the data a specific
two-morphism of bundle gerbes.
From such a Lagrangian description, equations
of motion can be derived \cite{rusu}. 

The present 
letter is a case study of defects in different exactly solvable CFT 
backgrounds which admit a geometric description as tori or toroidal
fibrations. For such backgrounds we use both the defect equations of
motion and standard techniques from D-branes
to make contact to structures encountered in 
geometric approaches to conformal field theory.

We explain the main tool used in Section 2 and 3: 
suppose that a defect $S$ separates  the worldsheet $\Sigma$ into two 
disjoint components $\Sigma_1$ and $\Sigma_2$. On $\Sigma_i$ the theory is
described by a target space $M_i$. In such a situation, the sigma model action
is defined for pairs of maps 
$X:\ \Sigma_1\to M$ and $Y:\ \Sigma_2\to \tilde M$.
The defect equations (\ref{dge}) for all cases considered in this paper
have schematically 
the form $\partial X=f_{\cal P}(\partial Y, \bar{\partial} Y)$
and $\bar{\partial} X=\bar{f}_{\cal P}(\partial Y, \bar{\partial} Y)$.
Given a boundary condition of the form $d(\partial X, \bar{\partial} X)=0$,
the defect equation  defines a new boundary condition
$d_{\cal P}(\partial Y, \bar{\partial} Y)=0$,
where $d_{\cal P}=d\circ (f_{\cal P},\; \bar{f}_{\cal P})$.
A notion of fusion between defects and a boundary can be expected
in the case of topological defects, since the latter can be moved to
the boundary without changing the correlator. For general defects, the
correlator depends on the relative position of the defect lines and a
notion of fusion is not obvious.

This reasoning provides in particular a direct relation
between T-duality of tori and Poincar\'e line 
bundles on the product of T-dual tori. Applying a
related reasoning to diagonal bibranes shows how
the action of a defect can result in a shift of the
B-field. Such a phenomenon is known from
monodromies in the moduli space of a Calabi-Yau 
compactification. In fact, the action of a defect
on a boundary condition can be expected \cite{Fuchs:2007fw}
to be of ``Fourier-Mukai'' type. This should be compared to Orlov's
theorem \cite{orlov} which provides a strong relation between fully 
faithful functors 
on the derived categories of coherent sheaves on smooth projective varieties 
and Fourier-Mukai functors. Inducing shifts of the B-field 
in terms of defects can be seen as a physical realization of such a
relation.

In a second part of this paper, we study T-dualities
in the special case of the group manifold $\SU(2)$ and
a lens space. In this study, we
use the fact that defects (at least
in rational conformal field theories) are uniquely determined by their
action on bulk fields. We construct several families of
defects by using T-duality and 
orbifold transformations. For one such family we determine
the geometry of the underlying bibranes. We recover 
structure familiar
from Fourier-Mukai transformations.
 
\section{Defect equations}

Let us consider a conformal interface between two conformal field
theories admitting a sigma model description with target spaces 
$M$ and $\tilde M$. For simplicity, we consider the situation described in the
introduction, where
the world sheet, a two-dimensional oriented manifold $\Sigma$, is separated by
an embedded oriented circle $S$ into two connected components. We take the 
convention that $\partial\Sigma_1=S$ and $\partial\Sigma_2=\bar{S}$ as 
equalities of oriented manifolds, where $\bar{S}$ is the manifold
$S$ with opposite orientation.

In such a situation, the sigma model action is defined for a pair
of maps
\be
X: \,  \Sigma_1\rightarrow M \,\,\ \mbox{and} \ \,\, 
Y: \,  \Sigma_2\rightarrow \tilde M \, . 
\ee

On the defect line $S$ itself, one has to impose conditions that relate
the two maps. The necessary data are captured by the geometrical
structure of a bibrane (for complete definitions see \cite{Fuchs:2007fw}): 
a bibrane is in particular
a submanifold of the Cartesian product of the target spaces, 
$Q\subset M\times \tilde M$. The pair of maps $(X,Y)$ is restricted by
the requirement that the combined map
\be \begin{array}{llll}
X_S: & S&\rightarrow & M\times \tilde M\\ 
&s&\mapsto & (X(s), Y(s))
\end{array}
\ee
takes its values in the submanifold $Q$.
Locally on each target space, the three-form field strength $H$ on
$M$ and $\tilde H$ on $\tilde M$ can be written as derivatives of
two-form connections $B$ and $\tilde B$:
\be H=\rmd B \quad \mbox{and}\quad \tilde H=\rmd\tilde{B} \,\, .  \ee
A second geometric datum of a bibrane is a gerbe bimodule on the bibrane
world volume $Q$. It contains as a particular piece of structure a globally
defined two-form $\omega$ on $Q$ such that locally 
the relation
\be
\label{onef}
\omega=p_1^*B-p_2^*\tilde{B}+\rmd A
\ee
with a local one-form $A$ holds. Equation (\ref{onef}) implies 
\be\label{h1h2}
d\omega=p_1^*H-p_2^*\tilde{H}
\ee

The Lagrangian then reads in terms of these local data:
\be
\label{defact}
S=\int_{\Sigma_1}L_1+\int_{\Sigma_2}L_2+\int_S X_S^{\ast}A
\ee
with the usual bulk Lagrangians
\be \begin{array}{lll}
L_1&=&G_{ij}\partial_{\alpha}X^i\partial^{\alpha}X^j+\epsilon^{\alpha\beta}B_{ij}\partial_{\alpha}X^i\partial_{\beta}X^j\\
L_2&=&\tilde{G}_{ij}\partial_{\alpha}Y^i\partial^{\alpha}Y^j+\epsilon^{\alpha\beta}\tilde{B}_{ij}\partial_{\alpha}Y^i\partial_{\beta}Y^j \,\, . 
\end{array}
\ee

From now on, we consider the conformal field theory
on the complex plane, parametrized by a complex
coordinate $z=\tau+\mathrm{i}\sigma$.
We take $\Sigma_1$ to be the left half-plane ($\sigma\leq 0$), and $\Sigma_2$ 
the right half plane ($\sigma\geq 0$) separated by the vertical
axis $S$, parametrized by $\tau$.
To write down the defect equations of motion, we assume 
that the world volume $Q$ of the bibrane is 
parametrised by coordinates $\eta^A$, 
$A=1,\ldots,k$ and functions $f^i$ with $i=1,\ldots, \dim M$
and $h^j$ with $j=1,\ldots, \dim \tilde M$:
\bea
&&X^i=f^i(\eta)\\ \nonumber
&&Y^j=h^j(\eta)
\eea
Then the defect equations of motion (see also equation (2.22) in \cite{rusu})
take following form:
\be
\label{dge}
\partial_A f^iG_{ij}\partial_{\sigma}X^j-\partial_A h^i\tilde{G}_{ij}\partial_{\sigma}Y^j
+(B_{ij}\partial_A f^i\partial_B f^j-\tilde{B}_{ij}\partial_A h^i\partial_B h^j+F_{AB})\partial_{\tau}\eta^B=0 \,\,\, ,
\ee
where $F=\rmd A$ is the two-form curvature of $A$.

As a side remark, we mention that a generalization of
the arguments of \cite{Abouelsaood:1986gd,Callan:1986bc,Leigh:1989jq} 
should allow to deduce that 
conformal invariance of the defect requires the fields to minimise a
modified Dirac-Born-Infeld action

\be
\label{dbd}
S_{DBI}=\int d\eta\; e^{\phi/2}\sqrt{\det(G_{AB}+\tilde{G}_{AB}+B_{AB}-\tilde{B}_{AB}+F_{AB})}
\ee
where $G_{AB}=G_{ij}\partial_A f^i\partial_B f^j$, $\tilde{G}_{AB}=\tilde{G}_{ij}\partial_A h^i\partial_B h^j$,
$B_{AB}=B_{ij}\partial_A f^i\partial_B f^j$ and $\tilde{B}_{AB}=\tilde{B}_{ij}\partial_A h^i\partial_B h^j$.

The Dirac-Born-Infeld action (\ref{dbd}) is consistent with the gauge invariance
\bea
\label{symd}
&&B\rightarrow B+d\Lambda\\ \nonumber
&&\tilde{B}\rightarrow \tilde{B}+d\tilde{\Lambda}\\ \nonumber
&&A\rightarrow A-\Lambda+\tilde{\Lambda}
\eea
of the world sheet action (\ref{defact}). The action (\ref{dbd}) should be compared to 
the ``folding 
trick'' \cite{Oshikawa:1996dj,Wong:1994pa,Bachas:2007td,Bachas:2001vj}
which describes defects in terms of D-branes on product 
$\rm{CFT}_1\times\overline{\rm{CFT}}_2$,
where $\overline{\rm{CFT}}_2$ denotes some
``conjugate'' conformal field theory, obtained by exchanging  
in particular left- and right movers. This is compatible with the
flipping of the sign of the antisymmetric B-field
in the modified Dirac-Born-Infeld action (\ref{dbd}).

\section{Toroidal backgrounds}

We now consider more specifically the case when both target
spaces are tori, $M=T$ and $\tilde M=\tilde T$. To obtain a $\sigma$-model
with conformal invariance, the gerbes are chosen to be 
topologically trivial with constant background metric $G$ and constant
B-field. They give constant background matrices
which have a symmetric and an antisymmetric
part, $E=G+B$ and $\tilde{E}=\tilde{G}+\tilde{B}$.

\subsection{T-dual tori}

We first consider bibranes that fill the whole product space,
$Q=T\times \tilde T$. Since the gerbes are assumed to be trivial, we have to consider
an ordinary line bundle on $Q$ which we take to be topologically trivial,
with connection given by a globally defined one-form
$A=A^X_i(X,Y)\rmd X^i+A^Y_i(X,Y)\rmd Y^i$.
The curvature two-form of this line bundle naturally 
decomposes into four parts $F^{XX}$, $F^{XY}$,
$F^{YX}$ and $F^{YY}$. Antisymmetry implies 
$F^{XY}=-F^{YX}$.

In terms of the background matrices, the defect 
equations (\ref{dge}) take the 
following simple form:
\begin{equation}\begin{array}{lll}
E\partial X-E^T\bar{\partial}X+
F^{XX}(\partial+\bar{\partial})X
+F^{XY}(\partial+\bar{\partial})Y&=&0 \\
\tilde{E}\partial Y-\tilde{E}^T\bar{\partial}Y
-F^{YY}(\partial+\bar{\partial})Y
+F^{XY}(\partial+\bar{\partial})X&=&0 \, .
\end{array}
\label{tordef}
\ee

In the case of T-dual tori, $\tilde{E}=E^{-1}$,
specific conformal interfaces are provided by  
Poincar\'e line bundles. They fill the whole product space and have 
curvature two-form
\be
F^{XX}=F^{YY}=0 \quad \mbox{ and } F^{XY}=I \,\, , 
\ee
where $I$ is unit matrix. 

Poincar\'e line bundles have a well-known relation to T-duality which
has become  apparent in the context of 
supersymmetric string backgrounds.
Here Poincar\'e line bundles
 enter in the transformation of Ramond-Ramond
tensor fields under T-duality \cite{Bergshoeff:1995as,Eyras:1998hn}:
indeed, it is a well-established fact that,
in the absence of a Neveu-Schwarz B-field, the associated
Fourier-Mukai kernel $\cal P$ gives rise to the following isomorphism
in K-theory \cite{Hori:1999me}
\be
T_{!}=\tilde{p}_!\circ \otimes {\cal P}\circ p^! : K^{\bullet}(T^n\times M)\rightarrow K^{\bullet+n}(\tilde{T}^n\times M)
\ee
where the tori $T$ and $\tilde T$ are T-dual and where we used the projections
$\tilde p: \tilde{T}^n\times T^n\times M\rightarrow \hat{T}^n$ and 
$p: \tilde{T}^n\times T^n\times M\rightarrow T^n$, respectively.

We now turn to study the bibranes given by Poincar\'e bundles in
purely bosonic conformal field theory: in this situation, the defect 
equations (\ref{tordef}) simplify:
\be\label{1}\begin{array}{lll}\label{2}
E\partial X-E^T\bar{\partial}X+\partial Y+\bar{\partial}Y
&=&0 \\
\tilde{E}\partial Y-\tilde{E}^T\bar{\partial}Y+
\partial X+\bar{\partial}X&=&0 \, .
\end{array}
\ee
Inserting the T-duality condition $\tilde{E}=E^{-1}$
in the second equation 
and multiplying by $E$ one finds
\be\label{4}
\partial Y-E(E^{-1})^T\bar{\partial}Y+E\partial X+E\bar{\partial}X=0 \, .
\ee
Inserting the first defect equation in (\ref{1}) in equation (\ref{4}) yields
\be\label{6}
-E(E^{-1})^T\bar{\partial}Y+E\bar{\partial}X+E^T\bar{\partial}X-\bar{\partial}Y=0 \,\, . 
\ee
Multiplication with the invertible matrix
$E^T(E+E^T)^{-1}$ gives
\be\label{8}
E^T\bar{\partial}X-\bar{\partial}Y=0 \, . 
\ee
The defect equations are thus equivalent to the two equations
\be
E\partial X=-\partial Y\\ \nonumber
\qquad E^T\bar{\partial}X=\bar{\partial}Y \, .
\ee
Transforming, as indicated in the introduction,
Neumann boundary conditions
\be E\partial X-E^T\bar{\partial}X=0 
\label{neumann} \ee
by such a defect yields Dirichlet boundary
conditions $(\partial+\bar{\partial})Y=0$. 
This is indeed the expected
action of T-duality in toroidal background, interchanging Dirichlet and Neumann boundary conditions.

\subsection{Shifts of the B-field}

We next consider a different situation: an identical 
toroidal background on both
sides of the defect, $T=\tilde T$ and $E=\tilde{E}$,
and defects described by bibranes 
supported on the diagonal $T\subset T\times T$. For a single 
compactified free boson,
this situation has been discussed in \cite{Fuchs:2007fw}.
Here, we focus on the new effects related to the presence of a non-trivial 
B-field on higher dimensional tori.

Our choice of bibrane imposes on all coordinates the equation
\be
\label{csd}
 \partial_{\tau}(X^i-Y^i)=0 \, .
 \ee
In the parametrization of the world volume $Q$ of the 
bibrane by the functions $X^i$, the defect equations
take the following form:
\be
\label{bedm}
E_{ij}\partial X^j-E^T_{ij}\bar{\partial} X^j-E_{ij}\partial Y^j+E^T_{ij}\bar{\partial} Y^j
+F_{ij}\partial_{\tau} X^j=0
\ee

The equations (\ref{csd}) and (\ref{bedm}) can be easily
seen to imply the two equations
\be \begin{array}{lll}
\partial X&=&\partial Y-{1\over 2}G^{-1}F(\partial Y+\bar{\partial} Y) \\
\bar{\partial} X&=&\bar{\partial} Y+{1\over 2}G^{-1}F(\partial Y+\bar{\partial} Y)
\label{sol2}
\end{array}
\ee
Substituting these expressions for $\partial X$ and 
$\bar{\partial} X$ in Neumann 
boundary conditions
\be
\label{nb}
(G+B)\partial X-(G-B)\bar{\partial}X=0
\ee
yields a shift of the B-field:
\be
\label{nbn}
(G+B-F)\partial Y-(G-B+F)\bar{\partial}Y=0 \, .
\ee

This shift in the B-field is the
simplest analogue of the following geometric fact that plays an
important role in superstring compactifications: tensoring
a complex of coherent sheaves ${\cal E}^{\bullet}$
on a complex variety $X$ with a line bundle $L$ on $X$, 
i.e.\
${\cal E}^{\bullet}\rightarrow {\cal E}^{\bullet}\otimes L$,
defines an autoequivalence of the bounded derived category of
coherent sheaves ${\cal D}^b(X)\rightarrow {\cal D}^b(X)$ which is 
just the Fourier-Mukai transform with kernel $\iota_*(L)$, where 
$\iota: X\rightarrow \Delta\subset X\times X$ is diagonal embedding of $X$.
The shift of the B-field also occurs for
Calabi-Yau varieties when one considers monodromies 
around divisors in moduli space where some even 
dimensional cycle vanishes 
\cite{Candelas:1990qd,Candelas:1994hw,Morrison:1995pf,Morrison:1995yi,Diaconescu:1999vp,brunner}.
In this case the monodromies amount to twisting with the 
line bundle associated to the divisor.

\section{Defects and T-duality on lens space}

\label{md}

The construction of the previous section can be applied fibrewise to torus
fibration and can be expected
 to relate
pairs of torus fibrations $\pi: \, E\to M$ and $\tilde\pi: \, \tilde E\to M$.
It should be noted that in this case the fibre product $E\times_M \tilde E$ 
appearing e.g.\ in \cite{Bouwknegt:2003vb,bunsch} can be identified with
the subset of elements of the product space $E\times \tilde E$ of
pairs $(e,\tilde e)$ with $\pi(e)=\tilde\pi(\tilde e)$. This submanifold is
the world volume $Q$ of the relevant bibrane.

We exemplify the situation in the case of the T-duality 
relating conformal sigma-models with
a lens space as a target space and the WZW theory
based on the compact connected Lie group $\SU(2)$ at level $k$.
The relevant lens space is a quotient of the group manifold
$\SU(2)$ by the right action of a cyclic
subgroup $\mathbb{Z}_k$ of a maximal torus
of $\SU(2)$. In Euler coordinates for $\SU(2)$, this corresponds 
to the identification
$\varphi\sim \varphi+{4\pi\over k}$. 

We use another parametrisation \cite{Giddings:1993wn} of the group
manifold $\SU(2)$ in terms of Pauli matrices,
\be
\label{par}
g=\mathrm{e}^{\mathrm{i}\phi{\sigma_3\over 2}}
\mathrm{e}^{\mathrm{i}\theta{\sigma_2\over 2}}
\mathrm{e}^{\mathrm{i}(\xi-\phi){\sigma_3\over 2}} \,\, , 
\ee
in which the metric takes form
\be
\rmd s^2=(\rmd\xi-(1-\cos\theta)\,\rmd\phi)^2+\rmd\theta^2+\sin^2\theta
\, \,\rmd\phi^2
\ee
with $\theta\in[0,\pi]$, $\phi\in[0,2\pi]$ and $\xi\in[0,4\pi]$.

Identifying $\xi$ as the fibre coordinate exhibits the structure of the group 
manifold $\SU(2)$ as an $S^1$ bundle over $S^2$ of monopole charge $1$, 
the Hopf bundle. Considering the 
bundle with charge $m$ amounts to the substitution
$\xi\rightarrow {\xi\over m}$.
Due to the orbifold description of the lens space
$\SU(2)/\Z_k$, the latter can be considered as an $S^1$-bundle over $S^2$ 
with Chern class $k$ and thus admits a parametrization as in (\ref{par}),
but with $\xi\sim \xi+{4\pi\over k}$.
It is convenient to reparameterize the lens space bundle coordinate 
$\tilde{\xi}$ as $\tilde{\xi}=\xi'/k$.

To the $S^1$-bundle description of the lens space
and the group manifold, we can apply the standard geometric
T-duality construction \cite{Bouwknegt:2003vb}
for torus fibrations. It involves a  
correspondence space $E\times_M\tilde{E}$,
where in our case $E:=\SU(2)$, $\tilde{E}$ is
the lens space and the base manifold is $M:=S^2$.
It  leads to the following relations for the first Chern classes of
the $S^1$-bundles on $M$ and the three-forms $H$ and $\tilde H$ on $E$ and
$\tilde{E}$, respectively: 
\be
F=c_1(E)=\tilde{\pi}_*\tilde{H}\;\;\; 
\mbox{and} \,\,\,\, \tilde{F}=c_1(\tilde{E})=\pi_*H\, ,
\ee
where $\pi_*$ is integration on the $S^1$-fibre. 
It is observed in \cite{Bouwknegt:2003vb} that the pullbacks 
$\pi^*F$ and $\tilde{\pi}^*\tilde{F}$ are exact on $E$ and $\tilde{E}$ 
respectively, and therefore can be written as
\be\label{exa}
\pi^*F=\rmd A\;\;\; \mbox{and}\;\;\; \tilde{\pi}^*\tilde F=\rmd\tilde A,
\ee where 
$A\in\Omega^1(E)$ and 
$\tilde A \in \Omega^1(\tilde E)$ are global one-forms on $E$ and
$\tilde E$, respectively, which are assumed to be normalized such that
\be
\pi_*A=1=\tilde{\pi}_*\tilde{A}
\ee
It is shown in Section 3 of \cite{Bouwknegt:2003vb}
that there exists a three-form $\Omega$ on the
base manifold $M$ that obeys the two relations
\be
\label{thf}
H=A\wedge \pi^*\tilde{F}-\pi^*\Omega\;\;\; {\rm and} \;\;\; 
\tilde{H}=\tilde\pi^*F\wedge\tilde{A} -\tilde{\pi}^*\Omega \,\, . 
\ee

One then introduces a two-form $\omega$ on the correspondence space 
$E\times_M\tilde{E}$ by
\be
\omega:=\tilde p^*\tilde A\wedge   p^*A
\ee
where $p$ and $\tilde{p}$ are the projections $E\times_M\tilde{E}\rightarrow E$ and $E\times_M\tilde{E}\rightarrow \tilde{E}$
respectively. They obey the relation  $\pi p=\tilde{\pi}\tilde{p}$.

It follows from (\ref{thf}) and (\ref{exa}) and commutativity $p^*\pi^*=\tilde{p}^*\tilde{\pi}^*$, that
\be\label{defeq}
\rmd\omega=-\tilde p^* \tilde H+p^*H
\ee

This two-form also enters \cite{Bouwknegt:2003vb} in the following
isomorphism of twisted cohomologies
\be\label{isom}
T_*: p_*\circ e^{\omega}\circ \tilde p^*: 
H^{\bullet}(\tilde E,\tilde H)\rightarrow H^{\bullet+1}(E,H) \,\,\, . 
\ee

Let us comment on the important role of the equation (\ref{defeq}) in the 
isomorphism (\ref{isom}).
It is shown in \cite{Bouwknegt:2003vb} that thanks 
to this equation $H$-twisted cohomologies mapped to $\tilde{H}$-twisted 
cohomologies. On the other hand,
this equation coincides with equation (\ref{h1h2}), which was derived in 
\cite{Fuchs:2007fw} from the requirement of a well-defined worldsheet action.
This coincidence can be seen as additional evidence for
the relation between defects and kernels of Fourier-Mukai transforms we propose
in this paper.

In the case when $E=\SU(2)$ and $\tilde E$ is a lens space,
this yields
\be\begin{array}{lll}
H&=&{k\over 16\pi^2}\sin\theta\ \rmd\phi  \ 
\rmd\theta \ \rmd\xi \\
F&=&{1\over 4\pi}\sin\theta\ \rmd\phi \ \rmd\theta \\
A&=&{1\over 4\pi}\left(\rmd\xi-(1-\cos\theta) \, \rmd\phi\right)
\end{array}
\ee
and
\be\begin{array}{lll}
\tilde{H}&=&{1\over 16\pi^2}\sin\theta \ \rmd\phi \ 
\rmd\theta \ \rmd\xi' \\
\tilde{A}&=&{1\over 4\pi}\left(\rmd\xi'-k(1-\cos\theta)
\ \rmd\phi\right) \\
\tilde{F}&=&{k\over 4\pi}\sin\theta \ \rmd\phi \ 
\rmd\theta
\end{array}
\ee
and thus surpressing the projectors $p$ and $\tilde p$ for brevity
in calculations in explicit coordinates
\be
\label{wed}
\tilde A\wedge A={1\over 16\pi^2}\left(\rmd a\wedge \rmd \xi+
(1-\cos\theta)\rmd\phi \wedge \rmd a\right) 
\ee
where $a$ is defined by the equation
\be
\tilde{\xi}={\xi'\over k}=\xi+{a\over k} \,\, .
\ee

\subsection{Defect operators on bulk fields}

In this section, we describe defects by their action
on bulk fields. In the case of rational conformal
field theories, it is known (see Proposition 2.8 of \cite{ffrs5}) that
this action characterizes a defect uniquely.

The bulk partition function for the rational conformal field
theory associated to a lens space is
\begin{equation}
\label{partlens}
Z(q)=\sum_{j=0}^{k/2} \sum_{n\in\mathbb{Z}}\,
\chi_j^{SU(2)}(q)\chi_{jn}^{PF}(\bar{q})\psi^{U(1)}_{-n}(\bar{q})
\, .
\end{equation}
To derive conformal defects between $\SU(2)_k$ and 
the lens space $\SU(2)/\Z_k$
we need the following endomorphisms of a direct sum
of Fock spaces for left movers and right movers, respectively:
\begin{eqnarray}
\label{ishstg}
P_{r\pm}^{U(1)} &=& \exp\left[
\pm\sum_{n=1}^{\infty}{\alpha_{-n}^0\alpha^1_{n}\over n}\right]
\sum_{l\in \Z}|{r+2kl\over \sqrt{2k}}\rangle_0 \,\, \otimes \,\, _1\langle\pm{r+2kl\over \sqrt{2k}}
| \, \\
\label{ishstgg}
\bar{P}_{r'\pm}^{U(1)} &=& \exp\left[
\pm\sum_{n=1}^{\infty}{\tilde{\alpha}_{-n}^0\tilde{\alpha}^1_{n}\over n}\right]
\sum_{l'\in \Z}
\overline{| \pm {r'+2kl'\over \sqrt{2k}}\rangle_0} \,\, \otimes\,\, 
_1\overline{\langle{r'+2kl' \over \sqrt{2k}}|}
  \, ,
\end{eqnarray}
where the subscripts $0$ and $1$ distinguish free boson theories on
the two sides of the defect. 
The bra- and ket-states are highest weight states in Fock spaces.
They obey the following conservation equations for the
$\U(1)$-currents 
\begin{equation}
\label{u(1)}
J_0^3\pm J_1^3=0 \, , \quad  \bar{J}_0^3\pm\bar{J}_1^3=0 \, ,
\end{equation}
where e.g.\ the first equation is a short hand for the intertwining property
$$ P_\pm^{U(1)} J_1^3 = \mp J_0^3 P_\pm^{U(1)} \,\,\,\, . $$
Similarly, we consider for the parafermion theories
${\cal A}_0^{PF(k)} \times {\cal A}_1^{PF(k)}$
the following two operators
\begin{eqnarray}
\label{ishfrtw}
P_{[j,n]}^{PF} &=&
\sum_{N}|j,n,N\rangle_0\, \otimes \, _1\langle j,n,N| \, , \\
\label{ishsetw}
\bar{P}_{[j,n]}^{PF} &=& 
\sum_{M}\overline{|j,n,M\rangle}_0 \, \otimes \, _1\overline{\langle j,n,M|} \, ,
\end{eqnarray}
where the sums over $M$ and $N$ are over orthonormal bases of the
parafermion state spaces.
Here $j \in \{0,\frac12,1,\ldots \frac k2\}$ and $n \in \Z/2k\Z$ satisfy the 
constraint $2j + n = 0\; {\rm mod}\; 2$. The pairs 
$(j,n)$ and $(k/2-j, k+n)$ have to be identified.

Our starting point are symmetry preserving defects
in the $\SU(2)$-theory.\ The corresponding operators
on bulk fields can be expressed in terms of the
modular matrix $S$ of $\SU(2)$  and the identity operators
on irreducible highest weight modules of the 
corresponding untwisted affine Lie algebra,
\begin{eqnarray}
\label{ishfrtwd}
P_j^{SU(2)} &=&
\sum_{N}|j,N\rangle_0 \, \otimes \, _1\langle j,N| \, , \\
  \bar{P}_j^{SU(2)}&=& 
\sum_{M}\overline{|j,M\rangle}_0 \,\otimes
\, _1\overline{\langle j,M|} \, ,
\end{eqnarray}
where the sums over $M$ and $N$ are over orthonormal bases of the
$\SU(2)$ state spaces.
These endomorphisms preserve, of course, all $\SU(2)$ symmetries,
\begin{eqnarray}
\label{bbc1a}
J_0^a  &+&  J_1^a=0 \, , \\
\label{bbc2a}
\bar{J}_0^a &+&  \bar{J}_1^a=0 \, , \quad (a=1,2,3) \, . 
\end{eqnarray}
The action of a symmetry preserving defect on bulk fields is given 
in terms of these endomorphisms by \cite{pezu5}:
\begin{equation}
\label{carsttw}
X_a=\sum_{j} \, {S_{aj}\,\ \,\over S_{0j}}P_j^{SU(2)}
\bar{P}_j^{SU(2)}
\, .
\end{equation}
Since in the situation at hand no field identification fixed points occur,
we  can apply the procedure described in 
\cite{Fuchs:2000fd,Maldacena:2001ky,Sarkissian:2003yw} 
to derive a new family of defects 
separating $\SU(2)_k$ and the lens space $\SU(2)/\Z_k$.
Performing a T-duality in (\ref{carsttw}) yields
\be
\label{y1}
Y_a^{AB}=\sum_{j}\sum_{n}{S_{aj}\over S_{0j}} \,\, \ P_j^{SU(2)}
\bar{P}_{j,n}^{PF}\bar{P}_{n-}^{U(1)}
\, .
\end{equation}
The defects (\ref{y1}) preserve all left moving currents, but only the right
moving current corresponding to the maximal torus,
\begin{eqnarray}
\label{bbc1b}
J_0^a  &+&  J_1^a=0 \, ,\quad (a=1,2,3)  \\
\label{bbc2b}
\bar{J}_0^3 &-&  \bar{J}_1^3=0 \, .  
\end{eqnarray}
As a consequence of these equations,
the defects (\ref{y1}) transform A-type branes on $\SU(2)_k$ to B-type brane on 
$\SU(2)/\Z_k$. 

A third family of defects is obtained by summing
over the images of (\ref{carsttw}) 
under the action of $\Z_k$, with a prefactor determined by the Cardy condition:
\be
\label{y2}
Y_a^{AA}= \sqrt k \, \sum_{j}{S_{aj}\over S_{0j}}P_j^{SU(2)}
\left(\bar{P}_{j,0}^{PF}\bar{P}_{0+}^{U(1)}+\bar{P}_{j,k}^{PF}\bar{P}_{k+}^{U(1)}\right)
\, .
\end{equation}
The defects (\ref{y2}) satisfy the conservation equations
\begin{eqnarray}
\label{bbc1c}
J_0^a  &+&  J_1^a=0 \, ,\quad (a=1,2,3)  \\
\label{bbc2c}
\bar{J}_0^3 &+&  \bar{J}_1^3=0 \, ,  \, . 
\end{eqnarray}
and transform A-type branes on $\SU(2)_k$ to A-type 
branes on the lens space $\SU(2)/\Z_k$. 

Performing a T-duality on the defects (\ref{y2}),
one derives another family of defects on 
$\SU(2)_k$ that map an A-type brane on $\SU(2)_k$ to a B-type brane on 
$\SU(2)_k$:
\be
\label{x2}
X_a^{AB}=\sqrt k\, \sum_{j}{S_{aj}\over S_{0j}}P_j^{SU(2)}
\left(\bar{P}_{j,0}^{PF}\bar{P}_{0-}^{U(1)}+\bar{P}_{j,k}^{PF}\bar{P}_{k-}^{U(1)}\right)
\, .
\end{equation}
The defects (\ref{x2}) satisfy the conservation equations
\begin{eqnarray}
\label{bbc1d}
J_0^a  &+&  J_1^a=0 \, ,\quad (a=1,2,3)  \\
\label{bbc2d}
\bar{J}_0^3 &-&  \bar{J}_1^3=0 \, ,  \, . 
\end{eqnarray}

Summing over images and performing T-duality in  the left moving sector of 
(\ref{y2}) yields a fifth family of defects that map B-type branes on 
$\SU(2)_k$ to A-type branes on $\SU(2)/\Z_k$:
\be
\label{y3}
Y_a^{BA}= k \sum_{j}{S_{aj}\over S_{0j}}
\left(P_{j,0}^{PF}P_{0-}^{U(1)}+P_{j,k}^{PF}P_{k-}^{U(1)}\right)
\left(\bar{P}_{j,0}^{PF}\bar{P}_{0+}^{U(1)}+\bar{P}_{j,k}^{PF}\bar{P}_{k+}^{U(1)}\right)
\, .
\ee
The defects (\ref{y3}) satisfy the conservation equations:
\begin{eqnarray}
\label{bbc1e}
J_0^3  &-&  J_1^3=0 \,  \\
\label{bbc2e}
\bar{J}_0^3 &+&  \bar{J}_1^3=0 \, ,  \, . 
\end{eqnarray}

\subsection{Geometry of defects}

We finally determine the geometry of the family of defects (\ref{y1})
relating $\SU(2)$ and the lens space $\SU(2)/\Z_k$. To this end,
we parametrize  bulk fields in terms of Euler angles $\vec\theta$ using
the representation function ${\cal D}_{m m'}^j$ of the spin $j$ 
representation:
\begin{equation}
\label{ishov}
|\vec{\theta} \rangle := \sum_{j,m,m'} \sqrt{2j+1} {\cal D}_{m m'}^j
  (\vec{\theta}) |j,m,m' \rangle \, .
\end{equation}
We are thus interested in the overlap
$\langle\vec{\theta}_0|Y_a^{AB}|\vec{\theta}_1\rangle$ as a function of two
sets of Euler angles. As in the calculation in \cite{Bordalo:2003fy}, 
the definition of the lens spaces as right quotients
implies that only terms of the defect operator (\ref{y1}) 
with $n=0,k$ contribute to the overlap; in the large~$k$ limit also the
term with $n=k$ can be ignored. Therefore,  we arrive in the limit
of large level $k$ at the function
\begin{equation}
\label{seccarov}
\langle\vec{\theta}_0|Y_a^{AB}|\vec{\theta}_1\rangle
\sim
\sum_{j}{k\over \pi}\sin[(2j+1)\hat{\psi}]\, 
{\cal D}^j_{00}(g_0^{-1}(\vec{\theta}_0)g_1(\vec{\theta}_1)) \,\, , 
\end{equation}
where the angle $\hat \psi$ is given in terms of $a$ by 
$\hat\psi:= \frac{(2a+1)\pi}{k+2}$.

To proceed, we express \cite{Maldacena:2001ky} the Wigner {\cal D}-functions 
in terms of Legendre polynomials as ${\cal
D}^j_{00}=P_j(\cos\theta)$. We find for the sum appearing on
the right hand side of equation (\ref{seccarov})
\begin{equation} \begin{array}{lll}
\sum_j \frac{\rme^{(2j+1)\rmi\hat\psi} - \rme^{-(2j+1)\rmi\hat\psi}}{2\rmi}
P_j(\cos\theta) 
&=& \frac{\rme^{\rmi\hat\psi}}{2\rmi}
\sum_j \rme^{(2j)\rmi\hat\psi} P_j(\cos\theta) 
- \frac{\rme^{-\rmi\hat\psi}}{2\rmi}
\sum_j \rme^{-(2j)\rmi\hat\psi} P_j(\cos\theta) 
\end{array} 
\end{equation}
which allows us to use the
generating function for Legendre polynomials
\begin{equation}
\label{genleg}
\sum_n t^n P_n(x)={1\over \sqrt{1-2tx+t^2}}\, .
\end{equation}
to simplify equation (\ref{seccarov}). We evaluate the sum on the right
hand side of equation (\ref{seccarov}) to
\begin{equation} 
\frac{\rme^{\rmi\hat\psi}}{2\rmi} \frac1{\sqrt{\rme^{2\rmi\hat\psi}
\left(\rme^{-2\rmi\hat\psi} - 2 \cos\theta + \rme^{2\rmi\hat\psi}\right)}}
+ \mbox{c.c.}  \\
= -\frac1{2\sqrt2} \frac1{\sqrt{\cos\theta - \cos\hat\psi}} 
+ \mbox{c.c.}    \,\, , 
\end{equation}
and thus the overlap to
\begin{equation}
\label{geom}
\langle\vec{\theta}_0|Y_a^{AB}|\vec{\theta}_1 \rangle
\sim
{\Theta(\cos\delta-\cos2\hat{\psi})\over
\sqrt{\cos\delta-\cos2\hat{\psi}}} \,\,\,\, . 
\end{equation}
Here $\Theta$ is the Heavyside step function and $\delta$ is the second Euler 
angle of the product element $g^{-1}_0(\vec{\theta}_0)g_1(\vec{\theta}_1)$.
Thus equation (\ref{geom}) implies \cite{Sarkissian:2002ie,Sarkissian:2002bg} 
that the ``difference'' $g^{-1}_0(\vec{\theta}_0)g_1(\vec{\theta}_1)$
takes its values in a subset consisting of products of an element in a
fixed conjugacy class $C$ with an element $L\in \U(1)$:
  \be
g^{-1}_0(\vec{\theta}_0)g_1(\vec{\theta}_1)\in CL \, . 
\ee

We next determine the two-form $\omega$ satisfying equation $(H_1-H_2)|_{\rm bibrane}=
\mathrm{d}\omega$ that is part of the bibrane-data.
Its value  in the element $xfx^{-1}L$ with $f$ a 
fixed element of the conjugacy class $C$ and
$x\in G$ and $L\in\U(1)$ arbitrary can be derived from the Polyakov-Wiegmann
identity
$$ \omega^{WZ}(gh) = \omega^{WZ}(g) + \omega^{WZ}(h) -
\rmd \mathrm{Tr}(g^{-1}\rmd g \,\, \rmd h h^{-1}) $$
for the Wess-Zumino three-form $\omega^{WZ}(g)={1\over 3}\mbox{Tr}(g^{-1}dg)^3$ as follows: we compute for
$g_0^{-1}g_1\in CL$ the difference
$$ \begin{array}{lll}
\omega^{WZ}(g_0) - \omega^{WZ}(g_1) &=&
\omega^{WZ}(g_0) - \omega^{WZ}(g_0 C L) \\[.3em]
&=&
\omega^{WZ}(g_0)  - [ \omega^{WZ}(g_0) + \omega^{WZ}(CL)
-\rmd {\rm Tr} \left(g_0^{-1}\rmd g_0 \rmd(CL)\,\, (CL)^{-1}\right) ] \\[.3em]
&=& - \omega^{WZ}(C) + \rmd{\rm Tr}\left(C^{-1}\rmd C \,\,\rmd L L^{-1} \right)
+ \rmd{\rm Tr} \left(g_0^{-1}\rmd g_0\,\, \rmd(CL) (CL)^{-1}\right) \, . 
\end{array} $$

As a consequence, the two-form 
\be
\omega:={k\over 8\pi^2}{\rm Tr}(C^{-1}\rmd C\, \rmd LL^{-1}
+g_0^{-1}\,\rmd g_0\, \rmd (CL)(CL)^{-1})-\omega^f(x) \,\,\,\, ,
\ee
where the two form
\be
\omega^f(x)={k\over 8\pi^2}{\rm Tr}(x^{-1}\rmd xfx^{-1}\, \rmd xf^{-1})
\ee
obeys $\rmd \omega^f(C) = {k\over 8\pi^2}\omega^{WZ}(C)$,
has the desired property ${k\over 8\pi^2}\omega^{WZ}(g_0) - {k\over 8\pi^2}\omega^{WZ}(g_1)=\rmd \omega$.
The coefficient fixed by the requirement
$\int_{SU(2)}{k\over 8\pi^2}\omega^{WZ}(g)=k$
to make contact with the geometrical consideration.

Asymptotically, for large $k$, the situation simplifies in the case when $f\approx e$, and 
 the bibrane worldvolume, i.e.\ the
correspondence space, consists of all pairs
of the form $(g_0, g_0L)$, with $g_0\in\SU(2)$ and $L\in \U(1)$.
The corresponding two-form takes the form
\be
\label{omt}
\omega={k\over 8\pi^2}{\rm Tr}(g_0^{-1}\, \rmd g_0\, \rmd LL^{-1})
\ee
In this case, the defect acts as an isomorphism on bulk fields, and we
thus expect a relation to T-duality. Indeed, we find in the parametrization 
(\ref{par})
\be
(g^{-1}\rmd g)_{11}=-(g^{-1}\rmd g)_{22}=\mathrm{i}{\rmd\xi\over 2}-
\mathrm{i}\rmd\phi{1-\cos\theta\over 2} \, . 
\ee
Writing $L=\mathrm{e}^{\rmi {a\sigma_3\over 2k}}$, we see that the
two-form (\ref{omt}) coincides with the two-form
(\ref{wed}) from the geometric approach. This nicely demonstrates how
geometric structure familiar from Fourier-Mukai transformations is
encoded in the algebraic data describing defects.

\vskip4em

\noindent {\bf Acknowledgements} \\[1pt]
We are grateful to  J\"urgen Fuchs and Ingo Runkel for useful discussions
and comments on the paper. \\
G.S.\ would like to thank King's College London for hospitality.
Both authors received partial support from the Collaborative Research Centre 
676 ``Particles, Strings and the Early Universe - the Structure of Matter and 
Space-Time''.

\end{document}